\documentclass[a4paper,10pt,twoside]{cpc-hepnp}

\usepackage{multicol}
\usepackage{graphicx}
\usepackage{booktabs}
\usepackage{amssymb,bm,mathrsfs,bbm,amscd}
\usepackage[tbtags]{amsmath}
\usepackage{lastpage}

\begin{document}

\title{CLAS+FROST: new generation of photoproduction experiments at
Jefferson Lab.}

\author{E. Pasyuk$^{1,2;1)}$\email{pasyuk@jlab.org}%
\\{\it for the CLAS Collaboration}
}
\maketitle

\address{%
1~Arizona State University, Tempe, AZ 85287, U.S.A.\\
2~Jefferson Laboratory, Newport News, VA 23606, U.S.A.\\
}

\begin{abstract}
A large part of the experimental program in Hall B of the Jefferson
Lab is dedicated to baryon spectroscopy.  Photoproduction experiments
are essential part of this program.  CEBAF Large Acceptance
Spectrometer (CLAS) and availability of circularly and linearly
polarized tagged photon beams provide unique conditions for this type
of experiments.  Recent addition of the Frozen Spin Target (FROST)
gives a remarkable opportunity to measure double and triple
polarization observables for different pseudo-scalar meson
photoproduction processes.  For the first time, a complete or nearly
complete experiment becomes possible and will allow model independent
extraction of the reaction amplitude.  An overview of the experiment
and its current status is presented.
\end{abstract}



\begin{multicols}{2}

\section{Introduction}
Among the most exciting and challenging topics in sub-nuclear physics
today is the study of the structure of the nucleon and its different
modes of excitation, the baryon resonances.  Initially, most of the
information on these excitations came primarily from partial wave
analysis of data from ${\pi}N$ scattering.  Recently, these data have
been supplemented by a large amount of information from pion electro-
and photoproduction experiments.  Yet, in spite of extensive studies
spanning decades, many of the baryon resonances are still not well
established and their parameters (i.e., mass, width, and couplings to
various decay modes) are poorly known.  Much of this is due to the
complexity of the nucleon resonance spectrum, with many broad,
overlapping resonances.  While traditional theoretical approaches have
highlighted a semi-empirical approach to understanding the process as
proceeding through a multitude of $s$-channel resonances, $t$-channel
processes, and non-resonant background, more recently attention has
turned to approaches based on the underlying constituent quarks.  An
extensive review of the quark models of baryon masses and decays can
be found in Ref. \cite{CR}.  Most recently lattice QCD is making
significant progress in calculations of baryon spectrum.  While these
quark approaches are more fundamental and hold great promise, all of
them predict many more resonances than have been observed, leading to
the so-called ``missing resonance'' problem.  One possible reason why
they were not observed because they may have small coupling to the
${\pi}N$.  At the same time they may have strong coupling to other
final states ${\eta}N$, $\eta^{\prime}N$, KY, $2{\pi}N$.x
\end{multicols}
\begin{center}
\tabcaption{Polarization observables in pseudo-scalar meson
photoproduction.  The entries in parentheses signify that the same
polarization observables also appear elsewhere in the table.}
\label{tbl:phprod}%
\renewcommand{\arraystretch}{1.0}
\begin{tabular}{c|cccc|ccc|cccc}
\hline
  Beam& \multicolumn{4}{c|}{Target} & \multicolumn{3}{c|}{Recoil} &
 \multicolumn{4}{c}{Target + Recoil} \\
\hline
\hline
 & $-$ & $-$ & $-$ & $-$ & $x'$ & $y'$ & $z'$ & $x'$ & $x'$ & $z'$ & $z'$ \\
 & $-$ & $x$ & $y$ & $z$ & $-$ & $-$ & $-$ & $x$ & $z$ & $x$ & $z$ \\
\hline
 unpolarized & $\sigma_0$ & $0$ & $T$ & $0$ & $0$ & $P$ & $0$ &
               $T_{x'}$ & \!$-L_{x'}$\! & $T_{z'}$ & $L_{z'}$ \\
 linear pol. & $\ -\Sigma\ $ & $\ H\ $ & \!$(-P)$\! &
               $-G$ & $\ O_{x'}$ & \!\!$(-T)$\!\! & $\ O_{z'}$ &
               \!$(-L_{z'})$\! & \!\!$(\,T_{z'}\,)$\!\! &
               \!$(-L_{x'})$\! & \!\!$(-T_{x'})$\!\!\\
 circular pol. & $\quad 0\quad$ & $F$ & $0$ & $-E$ &
                 $-C_{x'}$ & $0$ & $-C_{z'}$ &
                 $0$ & $0$ & $0$ & $0$ \\
\hline
\end{tabular}
\end{center}
\begin{multicols}{2}
There are three objects in pseudo-scalar meson photoproduction which
can be polarized: photon beam, target nucleon and recoil baryon.
There are 16 observables which can be measured.  Table
\ref{tbl:phprod} lists all of them in three groups: beam-target, beam
recoil and target-recoil.  The photoproduction reactions can be
described in terms of four complex amplitudes.  Therefore, in order to
be able to perform full reconstruction of the amplitude one need to
measure at least eight carefully chosen observables from at least two
different groups.  This implies requirement of having polarized beam,
polarized target and recoil polarimetry.  Hyperons have a remarkable
feature: their weak decay is self-analyzing.  This gift of nature
allows us to measure their polarization without polarimeter.

A large part of the experimental program of Jefferson Lab and CLAS in
particular is dedicated to baryon spectroscopy.  There were several
CLAS running periods with circularly and linearly polarized photon
beams.  High quality data for the cross sections of
$\pi^0$\cite{Dugger07}, $\pi^+$\cite{Dugger09}, $\eta$\cite{Dugger02},
$\eta^{\prime}$\cite{Dugger06} and kaon photoproduction were obtained.
In addition to the cross sections\cite{Bradford06} for $K^+\Lambda$
and $K^+\Sigma^0$ final states the polarization of hyperons
$P$\cite{McNabb04} and polarization transfer $C_x/C_z$ were
measured\cite{Bradford07}.  The beam asymmetry $\Sigma$ was measured
with linearly polarized beam.  Significant amount of data for cross
sections and beam asymmetries was also accumulated at ELSA, MAMI,
GRAAL and LEPS.  However, without double polarization measurement it
is still impossible to resolve all ambiguities in the reaction
amplitude.  Several experiments\cite{FROST} were proposed to measure
double polarization observables in all reaction channels $\pi^0p$,
$\pi^+n$, ${\eta}p$, $\eta^{\prime}p$, KY, $\pi^+\pi^- p$ with CLAS,
circularly and linearly polarized photons and longitudinally and
transversely polarized target.

\section{Experimental Hall-B}

Experimetal Hall-B at Jefferson Lab provides a unique set of
instruments for these experiments.  One instrument is the CLAS
\cite{CLAS}, a large acceptance spectrometer which allows detection of
particles in a wide range of $\theta$ and $\phi$.  The other
instrument is a broad-range photon tagging facility \cite{tag} with
the recent addition of the ability to produce linearly-polarized
photon beams through coherent bremsstrahlung.  The remaining component
essential for the double polarization experiments is a frozen-spin
polarized target (FROST) .

\subsection{Tagged photon beams}

The Hall B photon tagger \cite{tag} covers a range in photon energies
from 20 to 95\% of the incident electron beam energy.  Unpolarized,
circularly polarized and linearly polarized tagged photon beams are
presently available.

\subsubsection{Circularly-polarized photon beam}

With a polarized electron beam incident on the bremsstrahlung
radiator, a circularly polarized photon beam can be produced.  The
degree of circular polarization of the photon beam depends on the
ratio $k=E_{\gamma}/E_e$, and is given by \cite{Olsen}
\begin{equation}
   P_\odot = P_e \cdot \frac{4k - k^2}{4 - 4k + 3k^2}.
\end{equation}
The magnitude of $P_\odot$ ranges from 60\% to 99\% of the incident
electron beam polarization $P_e$ for photon energies $E_\gamma$
between 50\% and 95\% of the incident electron energy.  CEBAF
accelerator routinely delivers electron beam with polarization of 85\%
and higher.  An example of tagged circularly polarized photon spectrum
and its degree of polarization is shown in Fig. \ref{cbeam}
\begin{center}
\includegraphics[width=7cm]{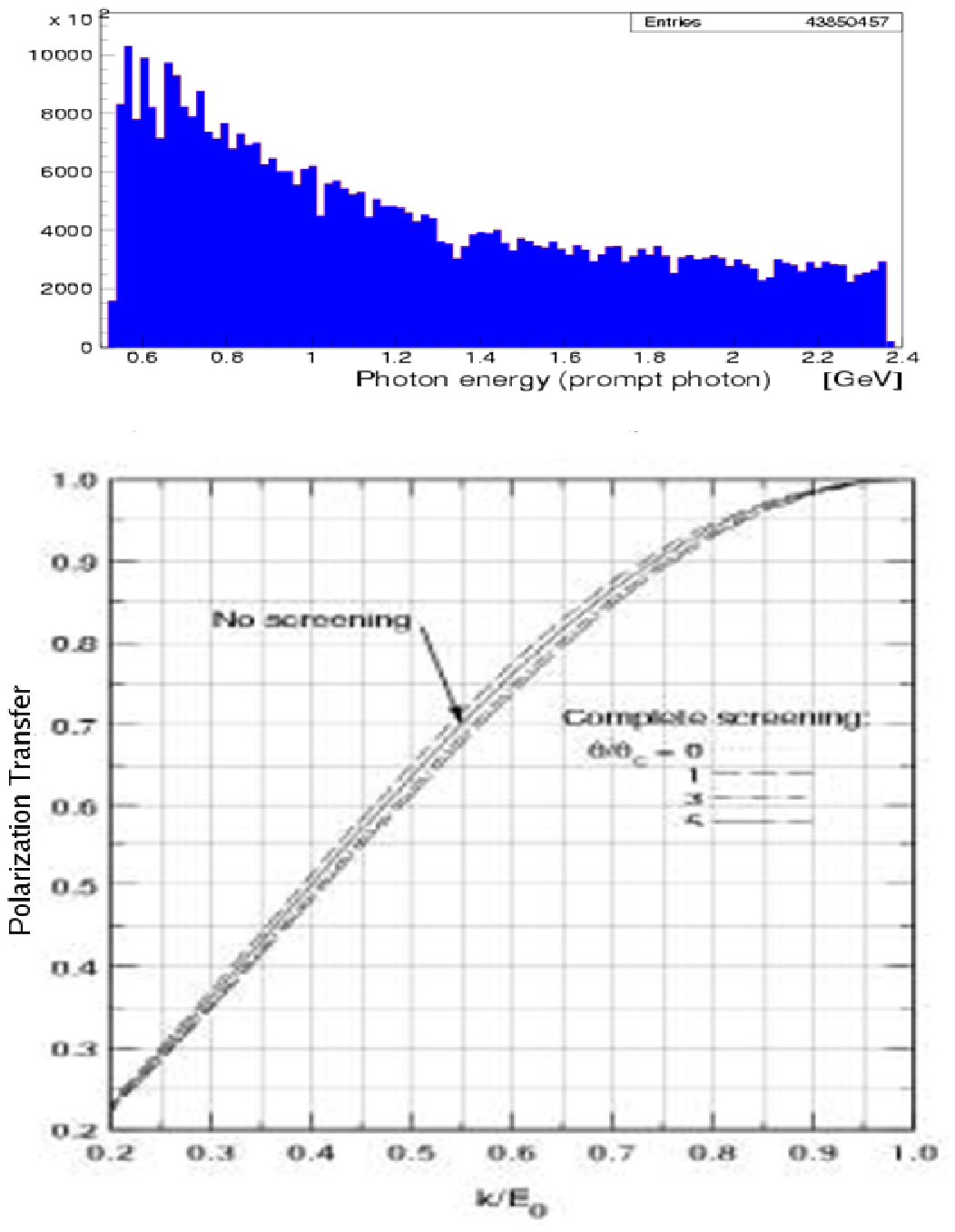} \figcaption{\label{cbeam} An
example of tagged bremsstrahlung photon spectrum (top) and
polarization transfer from electron to photon (bottom).  }
\end{center}

\subsubsection{Linearly polarized photon beam}

A linearly polarized photon beam is produced by the coherent
bremsstrahlung technique, using an oriented diamond crystal as a
radiator.  Figure \ref{lbeam} shows an example of collimated linearly
polarized tagged photon spectrum obtained in Hall-B.  The degree of
polarization is a function of the fractional photon beam energy and
collimation and can reach 80\% to 90\%.  With linearly polarized
photons, over 80\% of the photon flux is confined to a 200-MeV wide
energy interval.

\begin{center}
\includegraphics[width=7cm]{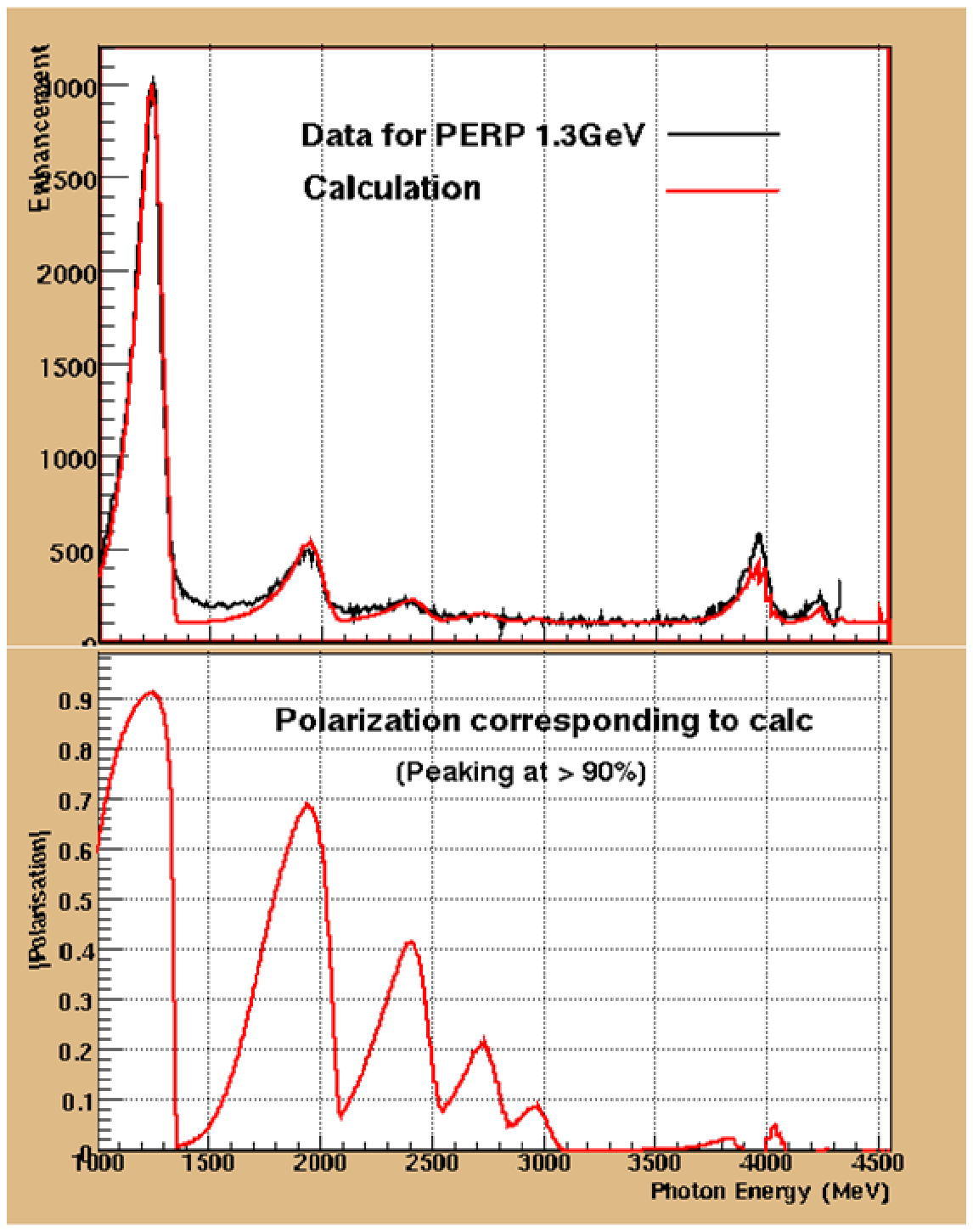} \figcaption{\label{lbeam}
Coherent bremsstrahlung spectrum and calculated photon polarization as
a function of energy}
\end{center}

\subsection{Frozen spin polarized target (FROST)}

An essential piece of the hardware for this experiment is a polarized
target capable of being polarized transversely and longitudinally with
a minimal amount of material in the path of outgoing charged
particles.  The Hall-B polarized target \cite{poltarg} used in
electron beam experiments is a dynamically polarized target.  The
target is longitudinally polarized with a pair of 5 Tesla Helmholtz
coils.  These massive coils limit available aperture to 55 degrees in
forward direction.  For photon beam experiments, a frozen-spin target
is a much more attractive choice.

In frozen-spin mode, the target material is dynamically polarized in a
strong magnetic field of 5 Tesla at the temperature of about 1K
outside of CLAS.  After maximal polarization is reached the cryostat
is turned to the ``holding'' mode with much lower magnetic field of
0.5 Tesla at a temperature of 50 mK or less, and then moved in CLAS.
A photon beam does not induce noticeable radiation damage and does not
produce significant heat load on the target material.  Under these
conditions the target can hold its polarization with a relaxation time
on the order of several days before re-polarization is required.
Since the holding field is relatively low, it is possible to design a
``transparent'' holding magnet with a minimal amount of material for
the charged particles to traverse on their way into CLAS.  The target
system uses an external polarizing magnet located outside of CLAS, and
an internal holding magnet inside the cryostat.  Butanol was chosen as
target material.  The operation cycle is following.
\begin{itemize}
\item Dynamically polarize target with microwaves outside of CLAS in
external polarizing magnet.
\item Turn off microwave and freeze polarization and switch to holding
mode with internal holding magnet.  Do spin rotation in case of
transverse polarization.
\item Take data for several days.
\item Repeat the cycle and flip polarization if required.
\end{itemize}

Figure \ref{target1} shows two configurations of the target.  On the
top panel the target is pulled out from the CLAS and inserted in the
polarizing magnet.  The bottom panels shows the target inside the CLAS
in it normal configuration for data taking.
\begin{center}
\includegraphics[width=7cm]{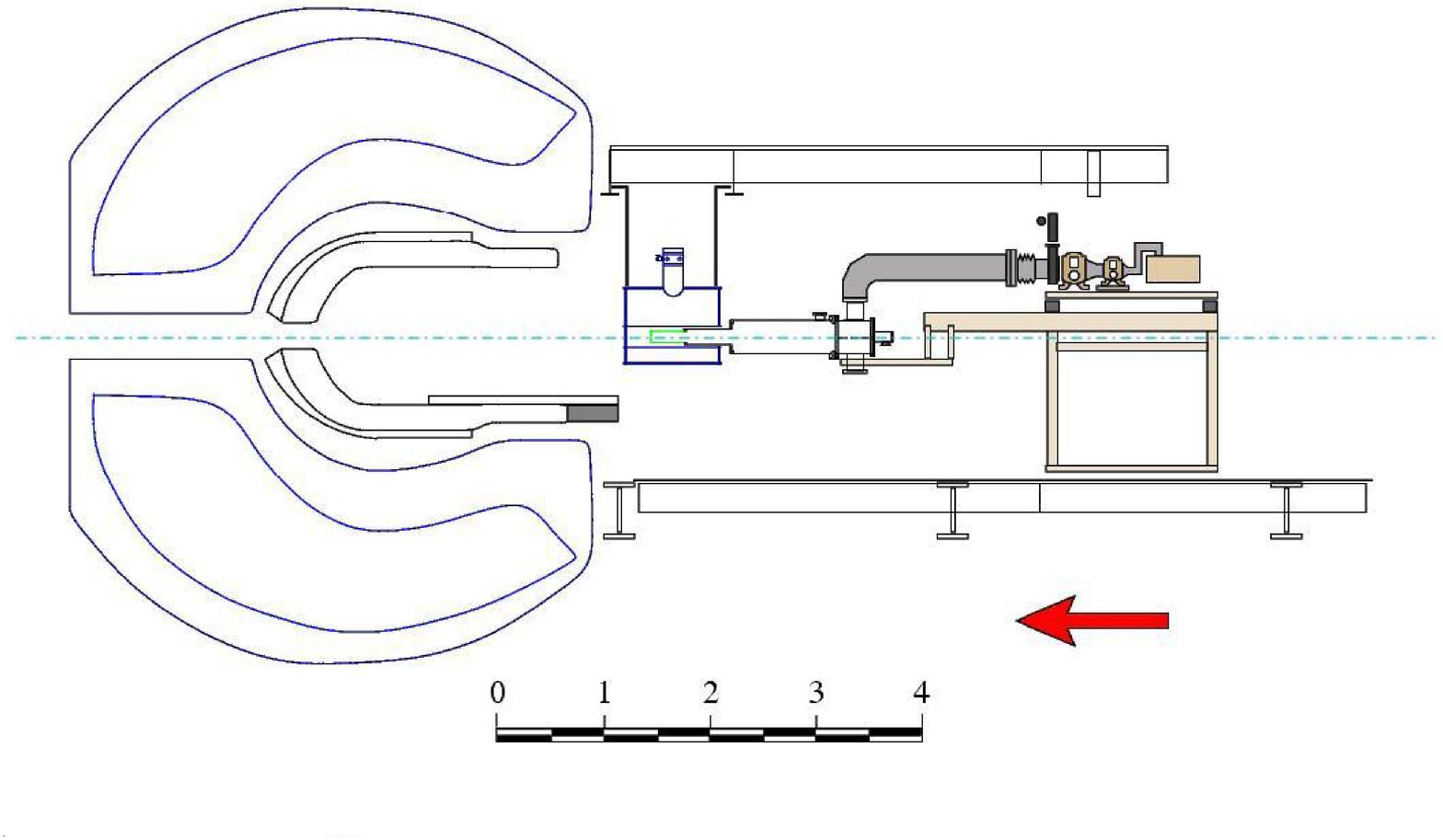}
\includegraphics[width=7cm]{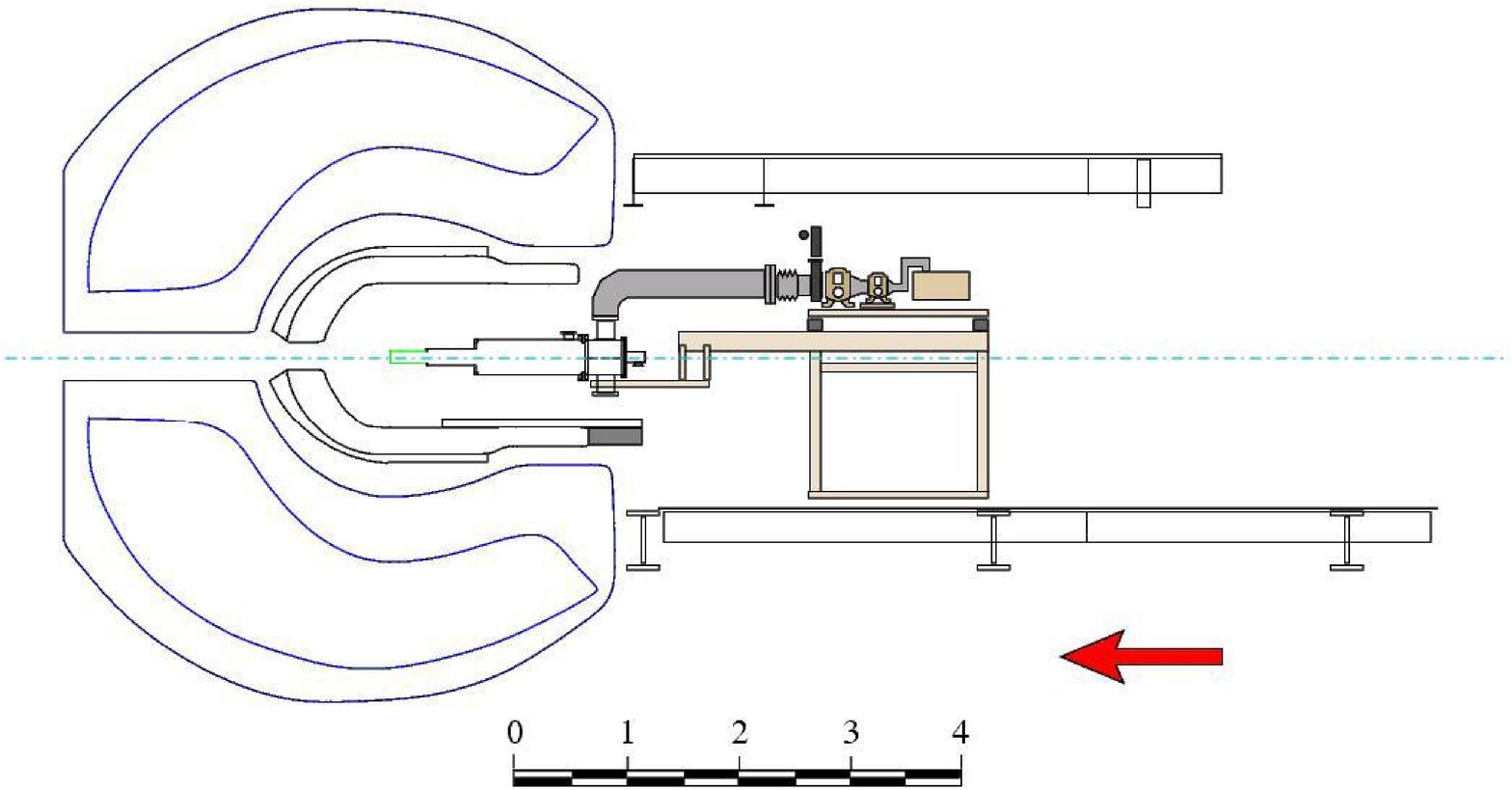} \figcaption{\label{target1}
The target is pulled out from the CLAS and inserted in the polarizing
magnet (top panel).  The target is inside the CLAS in its
normal configuration for data taking (bottom panel).  }
\end{center}

The frozen spin target has been built by the JLab polarized target
group.  Table \ref{targpar} summarizes main parameters of the target.
There are two holding magnets available an the target can be
configured either for longitudinal or transverse polarization.  During
the first round of experiments the target demonstrated excellent
performance running continuously for three and a half months.

\begin{center}
\tabcaption{Parameters of the FROST.}
\label{targpar}%
\begin{tabular}{|l|c|c|}
\hline
&Expectation&Result\\
\hline
Base         & 50 mK            & 28mK no beam \\
temperature  &                  & 30 mK beam \\
Cooling      & 10 $\mu$W        & 800 $\mu$W@50 mK \\
power        & (frozen)         & \\
             & 20 mW            & 60 mW@300 mK \\
             & (polarizing)     & \\
Polarization & $\pm85$\%        & +82\% \\
             &                  & -85\% \\
relaxation   & 500 hours        & 2700 h (+Pol) \\
time         & ($~$5\% per day) & 1400 h (-Pol) \\
             &                  & ($<$1.5\% per day) \\
\hline
\end{tabular}
\end{center}

\section{Experiment}

The first round of experiment with FROST was conducted in November
2007 -- February 2008.  In this set of experiments target polarization
was longitudinal. We used both, linearly and circularly polarized
photons.  Photon energy range covered 0.5 -- 2.3 GeV.  Trigger
required at least one charged particle in CLAS.  In addition to the
polarized butanol target we also had carbon and polyethylene target
downstream. They are useful for various systematics checks and for
determination of the shape of the background form bound nucleons in
butanol.  Target polarization was reversed at each re-polarization
cycle.  Combination of circularly polarized beam and longitudinally
polarized target allows us to measure helicity asymmetry E.  With
linearly polarized photons incident on longitudinal target we have
access to the observable G.

\section{Preview of the data}

In this section we present the first look at the data collected.  A
preliminary analysis of the helicity asymmetry in the reaction $\gamma
p \to \pi^+ n$ was done.  We used standard CLAS particle identification
procedure to select events with $\pi^+$ detected. For those events we
calculated missing mass assuming reaction $\gamma p \to \pi^+ X$.  The
missing particle was required to have a mass of neutron. Selected
events were binned in photon energy and $\cos \theta_{\rm c.m.}$.
In each bin we determined measured asymmetry
\begin{equation}
A = \frac{N_{\rightrightarrows}-N_{\rightleftarrows}}
{N_{\rightrightarrows} + N_{\rightleftarrows}},
\end{equation}
where $N_{\rightrightarrows}$ and $N_{\rightleftarrows}$ is number of
events beam and target polarization is parallel and anti-parallel
respectively. Then, to get helicity asymmetry, E, this raw asymmetry was
corrected for beam and target polarization and effective dilution
factor.  At this stage of the analysis we used very rough estimate of
these corrections A very preliminary results for helicity asymmetry
for $\pi^+$ photoproduction are presented in Figures \ref{figa} --
\ref{figd}.  This represents about 10\% of available statistics.

\begin{center}
\includegraphics[width=7cm]{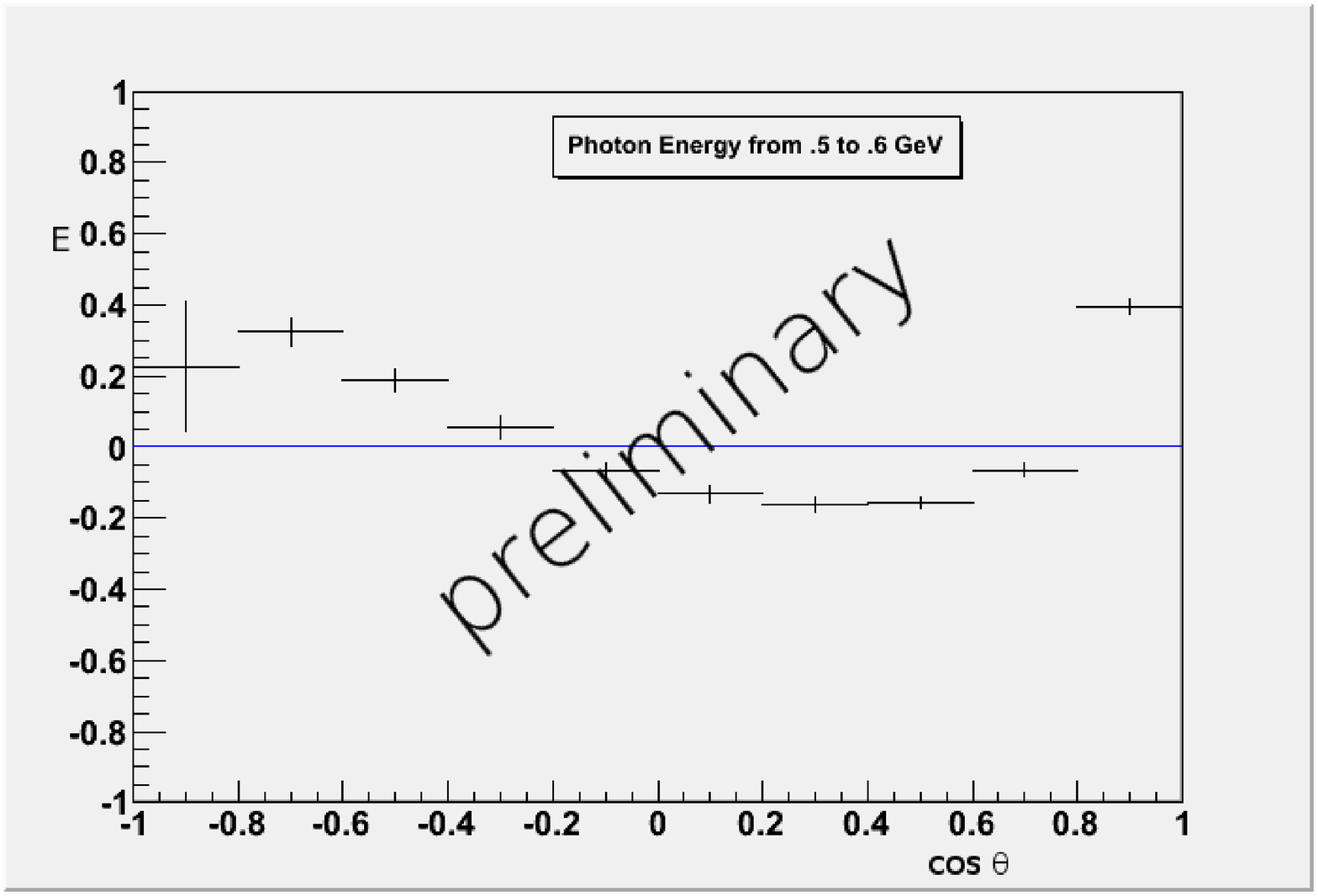} \figcaption{\label{figa}
Helicity asymmetry for $\gamma p \to \pi^+ n$. $E_\gamma = 0.5-0.6
GeV$. }
\vspace{3mm} \includegraphics[width=7cm]{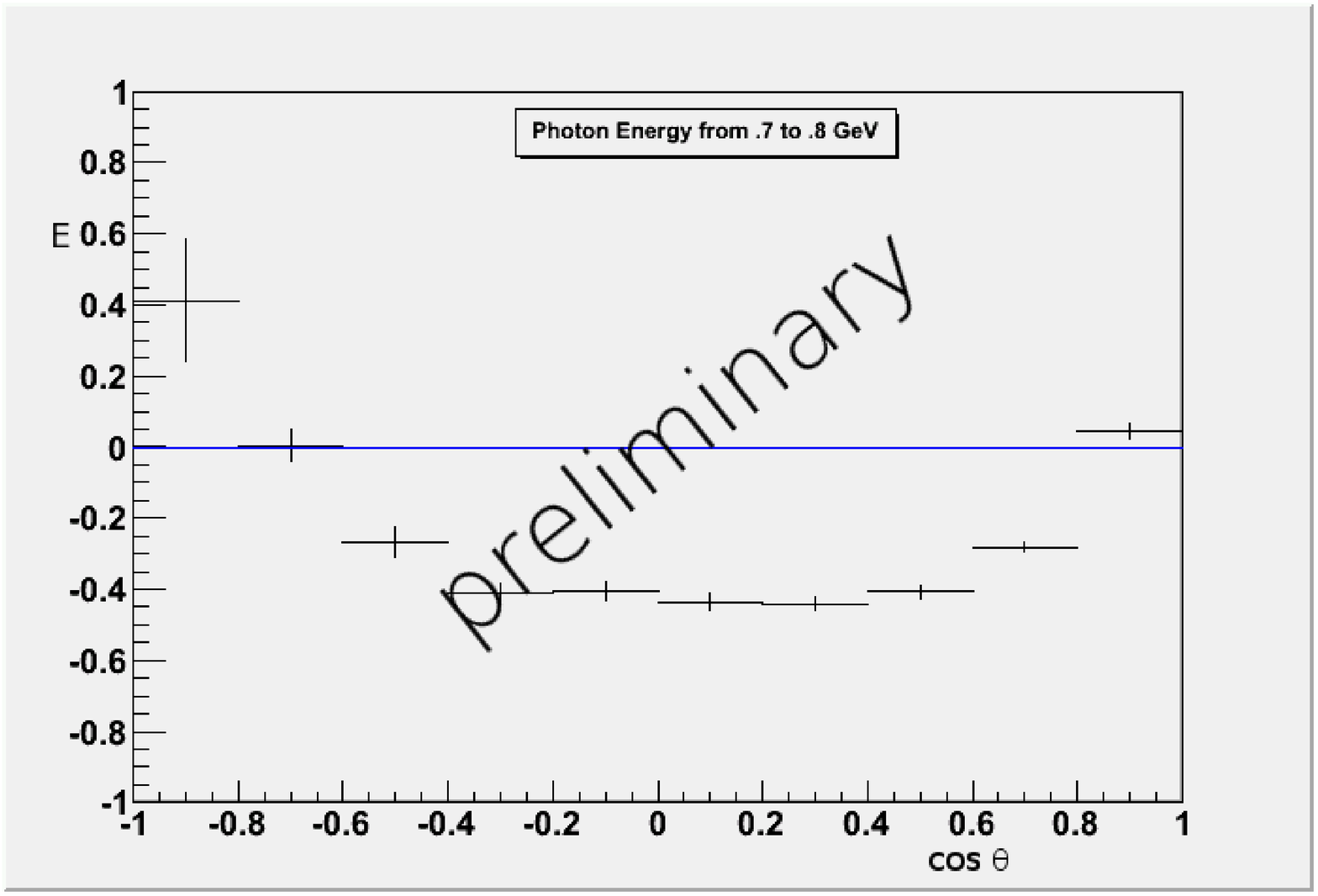}
\figcaption{\label{figb} Helicity asymmetry for $\gamma p \to \pi^+
n$. $E_\gamma = 0.7-0.8 GeV$. }
\vspace{3mm} \includegraphics[width=7cm]{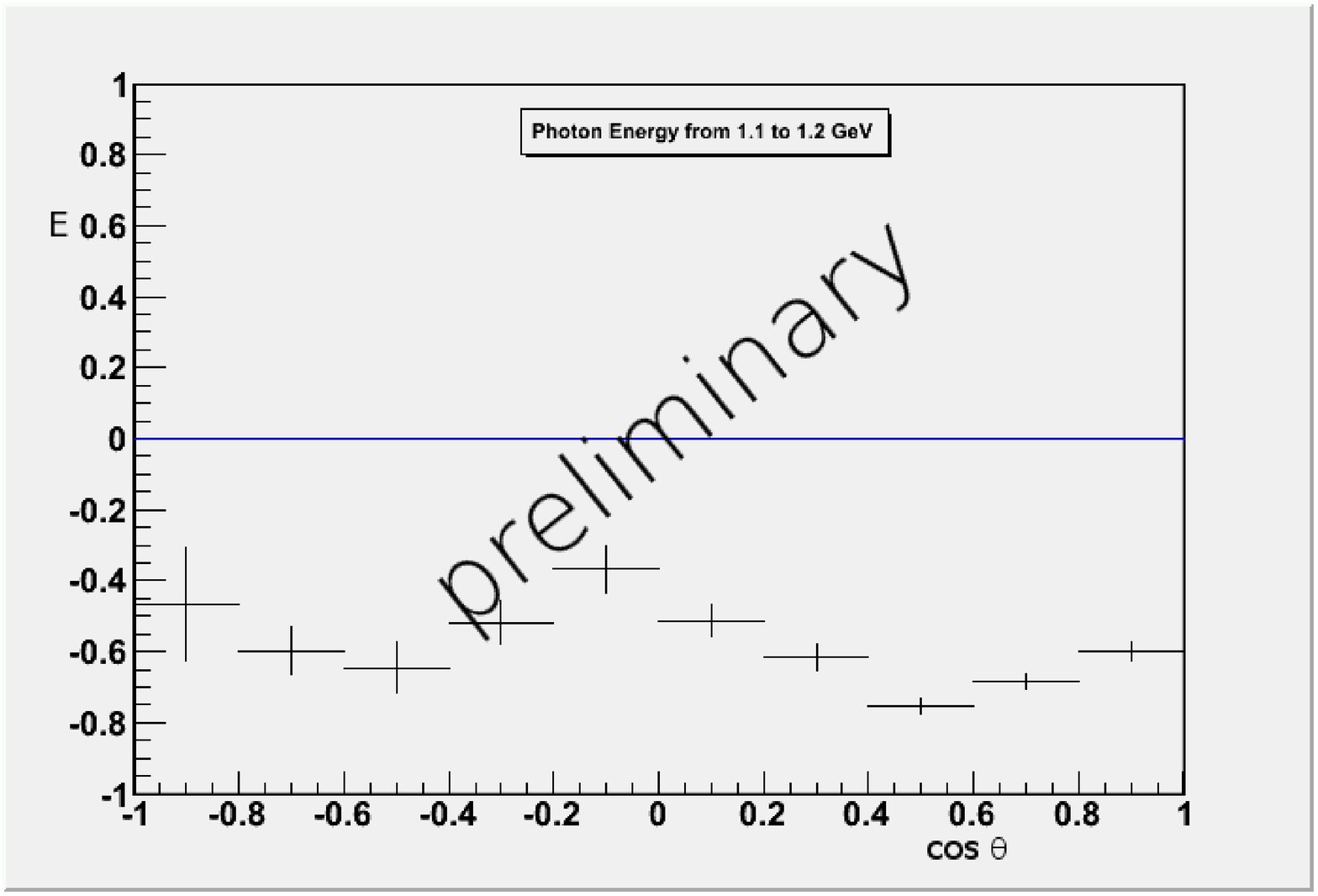}
\figcaption{\label{figc} Helicity asymmetry for $\gamma p \to \pi^+
n$. $E_\gamma = 1.1-1.2 GeV$. }
\vspace{3mm} \includegraphics[width=7cm]{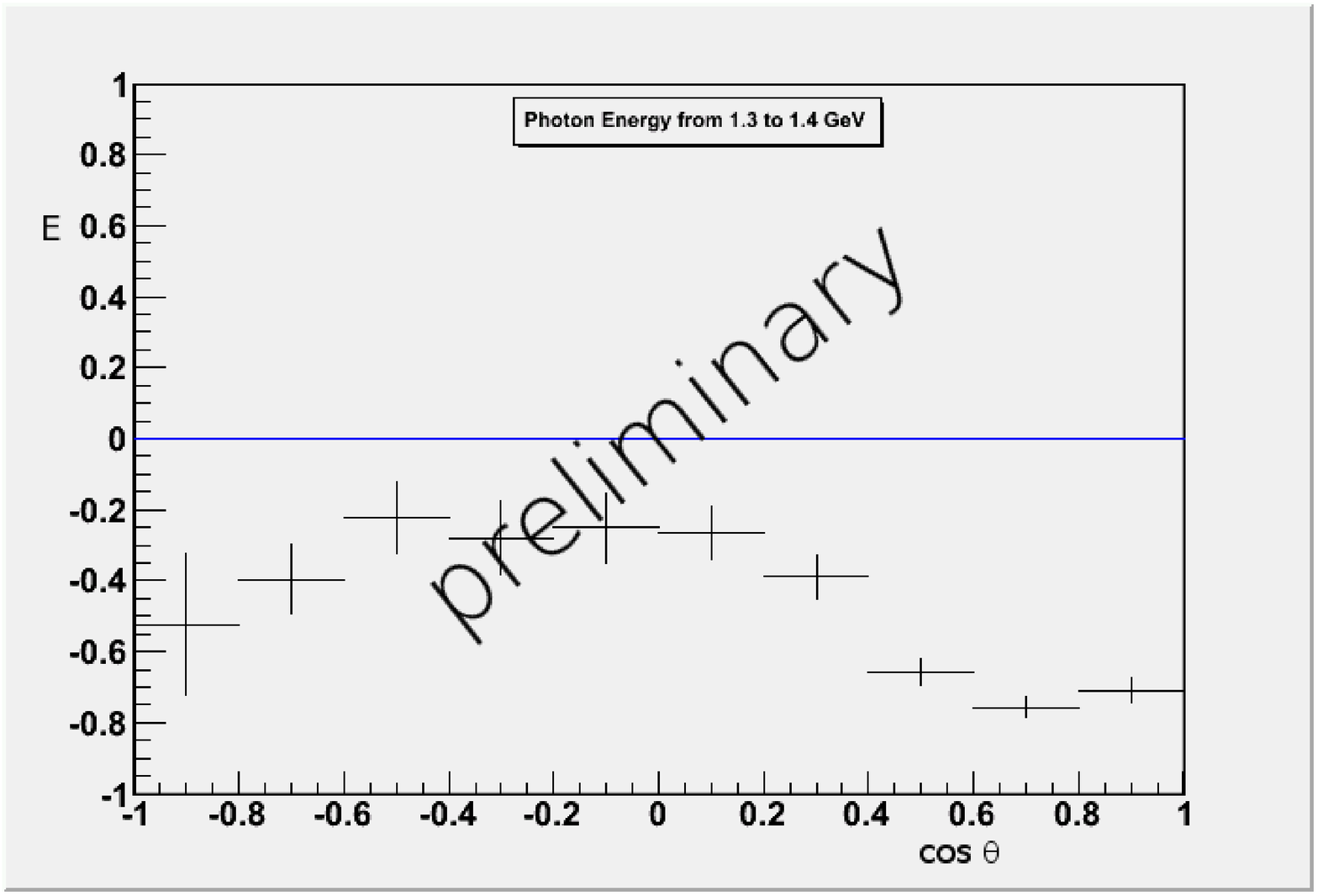}
\figcaption{\label{figd} Helicity asymmetry for $\gamma p \to \pi^+
n$. $E_\gamma = 1.3-1.4 GeV$. }
\end{center}

One can see that asymmetry is large and energy evolution displays
various structures in angular dependence.  This is just one example of
what is coming in near future.  The analysis of all other final
states is underway. 

\section{Summary}

The addition of Frozen Spin Target, with both, longitudinal and
transverse polarization significantly advances our experimental
capabilities.  The first round of the double polarization
photoproduction experiments with longitudinally polarized target has
been complete and experimental data are being analyzed.  The second
part of the experiment is scheduled to run in spring of 2010 and will
use transversely polarized target.

Upon completion of the experiment it will be possible for the first
time to perform complete experiment of KY photoproduction and nearly
complete for other final states.  Entire program is more than just a
sum of several experiments, observables for all final states are
measured simultaneously under the same experimental conditions and
have the same systematic uncertainties.  It can be considered as a
``coupled channel experiment'' ultimately providing data for coupled
channel analysis and extraction of parameters of baryon resonances.

Another essential part of the program which was not described here
involves photoproduction experiments on the deuteron target which
allow to study different isospin states of the baryon resonances.
Several CLAS experiments with polarized photons and unpolarized
deuteron target were complete.  Double polarization experiments with
HD-Ice polarized target are in preparation.

\acknowledgments{The authors gratefully acknowledge the work of the
Jefferson Lab Accelerator Division staff.  This work was supported by
the National Science Foundation, the U.S. Department of Energy (DOE),
the French Centre National de la Recherche Scientifique and
Commissariat \`a l'Energie Atomique, the Italian Istituto Nazionale di
Fisica Nucleare, and the Korean Science and Engineering Foundation.
The Southeastern Universities Research Association (SURA) operated
Jefferson Lab for DOE under contract DE-AC05-84ER40150 during this
work.}

\end{multicols}

\vspace{-2mm}
\centerline{\rule{80mm}{0.1pt}}
\vspace{2mm}

\begin{multicols}{2}

\end{multicols}

\end{document}